\documentclass[12pt,letterpaper]{article}
\pdfoutput=1
\usepackage{jheppub}
\usepackage{journals}
\usepackage{graphicx}

\def\Nfour	{\mathcal{N}\,{=}\,4}
\def\Nc		{N_{\rm c}}
\def\q		{\mathbf {q}}
\def\omegahat	{\hat\omega}
\def\qhat	{\hat q}
\def\half	{\frac {1}{2}}

\begin		{document}
\title		{Finite coupling corrections to holographic predictions
		for hot QCD}
\preprint	{HIP-2015-26/TH\\\hbox to \textwidth{\hfill INT-PUB-15-047}}

\author[a]	{Sebastian~Waeber,}
\author[a]	{Andreas~Sch\"afer,}
\author[b]	{Aleksi~Vuorinen,}
\author[c]	{and Laurence~G.~Yaffe}

\affiliation[a] {Institute for Theoretical Physics, University of Regensburg,
		D-93040 Regensburg, Germany}
\affiliation[b]	{Helsinki Institute of Physics and Department of Physics,
		P.O.~Box 64, FI-00014 University of Helsinki, Finland}
\affiliation[c]	{Department of Physics, University of Washington, Seattle,
		WA 98195, USA}
\emailAdd	{sebastian.waeber@physik.uni-regensburg.de}
\emailAdd	{andreas.schaefer@physik.uni-regensburg.de}
\emailAdd	{aleksi.vuorinen@helsinki.fi}
\emailAdd       {yaffe@phys.washington.edu}

\abstract
    {%
    Finite 't Hooft coupling corrections to multiple
    physical observables in strongly coupled  $\Nfour$
    supersymmetric Yang-Mills plasma are examined,
    in an attempt to assess the stability of the expansion in
    inverse powers of the 't Hooft coupling~$\lambda$.
    Observables considered include thermodynamic quantities,
    transport coefficients, and quasinormal mode frequencies.
    Although large $\lambda$ expansions for quasinormal mode
    frequencies are notably less well behaved than the expansions
    of other quantities, we find that a partial resummation of
    higher order corrections can significantly reduce the
    sensitivity of the results to the value of $\lambda$.
    }

\keywords	{holography and quark-gluon plasmas}
\maketitle

\section{Introduction}
\label{sec:intro}

Relativistic heavy ion collisions at the LHC and RHIC colliders produce a novel state of matter,
the quark gluon plasma (QGP) \cite{Shuryak:2004cy,Brambilla:2014jmp}.
At accessible energies, the produced plasma is strongly coupled --- so strongly coupled that
it cannot be described in terms of long-lived quasiparticles.
Evidence for this comes from the effectiveness of hydrodynamics
in modeling experimental results
(for a recent review, see ref.~\cite{Gale:2013da})
with viscosity close to the holographic value \cite{Kovtun:2004de},
as well as from the very short values of screening lengths in hot QCD
which are computable using lattice gauge theory
\cite{Laermann:2003cv}.

Hydrodynamics provides an effective description of the dynamics
of the plasma at sufficiently late times after the collision, but
is not applicable to early time dynamics when
the produced plasma is very far from local equilibrium.
Moreover, hydrodynamics is not adequate for understanding
the important physics of hard probes of the medium.
Unfortunately, alternative theoretical approaches for calculating,
reliably, properties of a strongly coupled plasma are very limited.

In recent years, gauge/gravity duality (or ``holography'') has provided
a new tool for understanding strongly coupled systems.
In its simplest and most studied form, gauge/gravity duality relates
properties of maximally supersymmetric $SU(\Nc)$ Yang-Mills
theory ($\Nfour$ SYM), in the $\Nc \to \infty$ limit,
to gravitational dynamics of higher dimensional asymptotically
anti-de Sitter spacetimes.
Under this duality, the process of equilibration and thermalization
in the quantum field theory is precisely related to gravitational dynamics
involving the formation and subsequent equilibration of black hole horizons.

Much work has been done using gauge/gravity duality to study aspects
of strongly coupled dynamics relevant to heavy ion collisions; see, for example,
refs.~\cite{Gubser:2009md,
	CasalderreySolana:2011us,
	Chesler:2015lsa}
and references therein.
This includes calculations of the drag on a heavy quark
    \cite{Herzog:2006gh,
	Caceres:2006as,
	Caceres:2006dj,
	Gubser:2006bz,
	CasalderreySolana:2007qw}
or light quark
    \cite{Chesler:2008uy,Gubser:2008as}
propagating through a strongly coupled medium,
jet quenching
    \cite{Chesler:2008wd,
	Chesler:2014jva},
particle production
\cite{CaronHuot:2006te,
	Hassanain:2012uj,
	Baier:2012ax,
	Baier:2012tc,
	Muller:2012rh},
isotropization dynamics
    \cite{Chesler:2008hg,
	Heller:2012km,
	Heller:2013oxa,
	Fuini:2015hba},
boost-invariant flow
    \cite{Chesler:2009cy,
    	Heller:2011ju},
collapsing bulk scalar fields, planar shells, and balls of dust
    \cite{Bantilan:2012vu,
	Bantilan:2014sra,
	Wu:2013qi,
	Danielsson:1999zt,
	Taanila:2015sda},
collisions of planar shock waves
    \cite{Chesler:2010bi,
	Casalderrey-Solana:2013aba,
	Casalderrey-Solana:2013sxa,
	Chesler:2015fpa},
and collisions of fully localized shock waves
resembling Lorentz contracted nuclei
    \cite{Chesler:2015wra, 
    Chesler:2015bba}.
Because of the precise mapping between gauge and gravitational
dynamics provided by gauge/gravity duality, in all this work
one is honestly computing properties of a strongly coupled
non-Abelian gauge theory.
There is just one problem --- the theory in which these calculations
are done, namely $\Nfour$ SYM, is not real QCD.

The strong coupling, large $\Nc$ limit of $\Nfour$ SYM, to which
gauge/gravity duality applies, may be viewed as a three-step deformation
of QCD:
(i) the fundamental representation quarks of QCD are replaced by
a collection of adjoint representation matter fields, both fermions
and scalars, thereby turning QCD into $\Nfour$ SYM,
(ii) the 't Hooft coupling $\lambda \equiv g_{\rm YM}^2 \, \Nc$,
which no longer runs with energy scale in $\Nfour$ SYM,
is tuned to very large values,
and (iii) the gauge group rank, $\Nc$, is sent to infinity. Qualitative properties of the deconfined plasma phase are stable under
these deformations:
the high temperature phase of the theory remains a non-Abelian plasma
with Debye screening and a finite correlation length;
spacelike Wilson loops continue to show area law behavior;
and long distance, low frequency dynamics continues to be described by
neutral fluid hydrodynamics.

Lattice gauge theory simulations have shown
that thermodynamic properties of $SU(\Nc)$ Yang-Mills plasma scale
very smoothly with $\Nc$ \cite{Panero:2009tv,Bali:2013kia},
suggesting that the large $\Nc$ limit
should be well-behaved for most observables of interest,
and moreover that the $SU(3)$ theory is already fairly close
to the $\Nc = \infty$ asymptotic limit.
Where results from hot QCD lattice simulations 
(in the experimentally relevant temperature range,
$1.5 \lesssim T/T_{\rm c} \lesssim 4$)
are available to be compared to holographic computations in $\Nfour$ SYM,
a variety of important physical quantities
such as the equation of state,
ratios of screening masses to temperature in various symmetry channels,
and estimates of the shear viscosity to entropy density ratio, $\eta/s$,
show agreement to within at least a factor of two, and often much better.

Consequently, in the above deformations which connect holographic
models to QCD, the step which likely produces the largest changes
in thermal properties, and about which the least is known,
is step (ii): sending the 't Hooft coupling to values large compared to unity.
At the relevant energy scales in hot QCD, the appropriate value of the
't Hooft coupling (in physically sensible schemes)
is presumably somewhere in the range 10--40 ---
corresponding to $\alpha_{\rm s} \equiv g^2_{\rm YM}/(4\pi)$ between 0.3 and 1,
not some truly enormous number.
Therefore, improved understanding of the dependence of physical quantities
in $\Nfour$ SYM on the value of $\lambda$ is highly desirable.
It is known that finite $\lambda$ corrections appear in the form of
inverse fractional powers, beginning with $\lambda^{-3/2}$.
In this paper, we collect, extend, and examine available results 
for finite-$\lambda$ corrections to thermal observables
in an effort to gain some insight into the stability of the expansion
in inverse powers of $\lambda$ and the applicability of holographic
predictions to physics at realistic values of the 't Hooft coupling.

The paper is organized as follows:
In section \ref{sec:results},
we summarize and discuss first order finite-$\lambda$ corrections
to a variety of thermal observables.
A basic observation is that the relative size of the first
finite-$\lambda$ correction is substantially larger for quasinormal
mode (QNM) frequencies than for other observables.
Section \ref{sec:correlators} then recaps the holographic calculation
of two point correlation functions,
from which transport coefficients and quasinormal mode frequencies
are extracted,
in a manner which allows one to extract the first order finite-$\lambda$
correction or perform a partial resummation of higher order finite-$\lambda$ corrections.
Results of the two procedures are shown, and compared, for the first
few quasinormal modes of the current-current correlator and the shear
channel of the stress-energy correlator,
as well as for the plasma conductivity and shear viscosity.
The final section \ref{sec:conclusion} contains a few concluding remarks.

\section {First order corrections: collected results}
\label{sec:results}

Considerable prior work exists examining finite-$\lambda$ corrections
to holographic results.
This includes analyses of
the equation of state \cite{Gubser:1998nz},
shear viscosity $\eta$ \cite{Buchel:2005,Buchel:2008sh},
plasma conductivity $\sigma$ \cite{Hassanain:2011fn},
photon production and transport \cite{Hassanain:2011ce,Yang:2015bva},
and various higher-order transport coefficients \cite{Grozdanov:2014kva}.
More recent work has also considered finite-$\lambda$ corrections to
quasinormal mode frequencies and off-equilibrium spectral densities
obtained from the current-current and stress-energy correlators
\cite{Steineder:2012si,Vuorinen:2013,Stricker:2013lma}.

The finite coupling corrections appear as a power series in $\lambda^{-1/2}$,
with the first corrections being proportional to $\lambda^{-3/2}$.
It proves convenient to define the constant
\begin{equation}
    \gamma
    \equiv \tfrac 18 \, \zeta(3) \, \lambda^{-3/2} 
    = \tfrac 18 \, \zeta(3) \, (g^2_{\rm YM} \Nc)^{-3/2} \,,
\end{equation}
where $\frac 18 \, \zeta(3) \approx 0.15$.
The benchmark range of 10--40 for the 't Hooft coupling $\lambda$
corresponds to values of $\gamma$ between about 0.005 and 0.0006.

\begin{table}[t]
%
\begin{center}

\begin{tabular}{|c|c|c|c|}
\hline
    Quantity &
    $\mathcal{O}(\gamma^0)$ &
    $\mathcal{O}(\gamma^1)$ &
    Reference
\\ \hline
    $s \, (\tfrac 12 \pi^2 \Nc^2 \, T^3)^{-1}$ &
    $1$ &
    $15 \, \gamma$ &
    \cite{Gubser:1998nz}
\\ \hline
    $\eta \, (\tfrac 18 \pi \Nc^2 \, T^3)^{-1}$ &
    $1$ &
    $135 \, \gamma$ &
    \cite{Buchel:2008sh}
\\ \hline
    $4\pi \, \eta/s$ &
    $1$ &
    $120 \, \gamma$ &
    \cite{Buchel:2008sh}
\\ \hline
    $\sigma \, (\tfrac 14 \alpha_{\rm EM} N^2 \, T)^{-1 }$ &
    1 &
    $14993/9 \, \gamma $&
    \cite{Hassanain:2011fn}
\\ \hline
    $\Gamma_0 \, (\alpha_{\rm EM} N^2 T^4)^{-1 }$ &
    0.053678 &
    $23.5379 \, \gamma $&
    This work
\\ \hline
    $\Gamma_1 \, (\alpha_{\rm EM} N^2 T^6)^{-1 }$ &
    0.472771 &
    $-224.4698 \, \gamma $&
    This work
\\ \hline
    ${\omega}^{\rm shear}_2 \, (2 \pi T)^{-1}$  &
    $2.585-2.382 \, i $ &
    $(1.029+0.957 \, i) \, 10^4 \, \gamma$  &
    \cite{Stricker:2013lma}
\\ \hline
    ${\omega}^{\text{EM}}_2 \, (2 \pi T)^{-1}$  &
    $2-2 \, i $ &
    $(1.34 +0.43 \, i) \, 10^5 \, \gamma$  &
    \cite{Vuorinen:2013}
\\ \hline
\end{tabular}
\end{center}
\caption
    {%
    \small
    Zeroth and first order terms in the expansion of
    various thermal observables in powers of
    $\gamma=\frac 18\, \zeta(3)\lambda^{-3/2}$.
    Results are shown for the entropy density $s$,
    shear viscosity $\eta$,
    viscosity to entropy density ratio $\eta/s$,
    electrical conductivity $\sigma$,
    the first two moments, $\Gamma_0$ and $\Gamma_1$,
    of the photoemission spectrum,
    and the second quasinormal mode frequencies,
    $\omega^{\rm EM}_2$ and $\omega^{\rm shear}_2$,
    at zero wavevector, for the electromagnetic current
    and shear channel of the stress-energy correlator,
    respectively.
   \label{table:1}
    }
\end{table}

In table \ref{table:1}, we collect the values of the first two terms
in the strong coupling expansions of
various thermal observables:
the entropy density $s$,
the shear viscosity $\eta$
and viscosity to entropy density ratio $\eta/s$,
the electrical conductivity $\sigma$,%
\footnote
    {%
    To define the SYM electromagnetic current and associated electrical conductivity,
    one weakly gauges a $U(1)$ subgroup of the global $SU(4)_R$ flavor symmetry
    of $\Nfour$ SYM
    \cite{CaronHuot:2006te};
    see the next section for details.
    }
and the second quasinormal mode frequencies $\omega_2^{\rm EM}$
and $\omega_2^{\rm shear}$
for the electromagnetic current and the 
shear channel of the stress-energy correlator, respectively.
Also included are results for the first two moments of the photoemission
spectrum, defined as
\begin{equation}
    \Gamma_n \equiv
    \int_0^\infty {dk} \> k^{2n} \,
    \frac{d\Gamma_\gamma}{dk} \,.
\label{eq:Gamma_n}
\end{equation}

For the entropy density (or equivalently, the pressure $p = s \, T/4$),
the first finite-coupling correction is modest;
the $O(\gamma^1)$ term does not exceed the leading $O(\gamma^0)$ term
as long as $\lambda > 1.72$.
For the shear viscosity or viscosity ratio $\eta/s$, the corresponding
crossover points where the first corrections equal the leading term
occur at $\lambda \approx 7.4$ or 6.9, respectively.
For the electric conductivity, this crossover lies at $\lambda \approx 39.7$,
while the crossovers for the photoemission moments $\Gamma_0$ and $\Gamma_1$
are at 16.3 and 17.2, respectively.
All these values are below, or at least within, our 10--40 range of benchmark
values for $\lambda$.
The situation, however, is rather different for the quasinormal
mode frequencies shown in table \ref{table:1}.
The first order corrections exceed
the leading order term when
$\lambda < 71.2$ (for $\omega_2^{\rm shear})$ or
$\lambda < 382.2$ (for $\omega_2^{\rm EM})$,
suggesting that their $\lambda = \infty$ limits are likely to give poor predictions
for the values of these quantities in the phenomenologically interesting
range of 't Hooft couplings.

A priori, it is not clear whether the above behavior of the QNM frequencies
is due to an abnormally large first term in an otherwise well-behaved expansion,
or whether the quantities in question are particularly sensitive to
finite coupling corrections, so that their expansions in $\gamma$ have
an abnormally small range of utility.
Deciding between these alternatives would,
in principle, require an all orders determination of the strong coupling
expansion, which is far beyond the reach of present day technology.
In this paper, we adopt a far more modest goal.
We will investigate
a simple resummation applicable to quasinormal mode frequencies
and related observables
which takes into account a subset of higher order
terms in the expansion in powers of $\gamma$,
and see if this improves the behavior of the resulting series.
At the very least, this investigation should be helpful in inferring
the range of utility of the above first order results.

\section{Finite coupling corrections to correlators and QNMs}
\label{sec:correlators}

Finite coupling corrections to thermal observables are generated by
higher derivative (or $\alpha'$) corrections to the 10-dimensional
type IIB supergravity action, which takes the schematic form
\begin{equation}
    S_{\rm IIB} =
    S_{\rm IIB}^{(0)}
    + \gamma \, S_{\rm IIB}^{(1)}
    + \gamma^{4/3} \, S_{\rm IIB}^{(4/3)}+ \cdots \,.
\label{eq_1}
\end{equation}
The first order correction $S_{\rm IIB}^{(1)}$ includes fourth powers
of the Riemann tensor plus terms, related by supersymmetry,
that involve the self-dual five form
(see, for example, ref.~\cite{Paulos:2008tn}).
For the free energy (or pressure), it is sufficient to evaluate the
first order correction terms in the action on the unmodified
AdS-Schwarzschild solution \cite{Gubser:1998nz}.
For other observables, one must insert into the 10D action
(\ref{eq_1}) an appropriate ansatz for the 10D metric,
five form field strength, and any other fields relevant
for the observable of interest.
A Kaluza-Klein reduction eliminating the compact internal space
(for physics which only depends on the lowest KK modes)
leads to a 5D $\alpha'$-corrected action for the relevant
bulk fields in asymptotically AdS spacetime.
Using the corrected 5D action, observables of interest
are computed using the standard holographic correspondence.
This typically involves deriving $\alpha'$ corrected
equations of motion for the relevant bulk fields
and then solving these equations order by order in $\gamma$.
For more details of such calculations
see, for example,~refs.~\cite{Gubser:1998nz,
	Buchel:2008sh,
	Myers:2008yi,
	Hassanain:2011fn,
	Vuorinen:2013}.

Below, we illustrate explicitly the above procedure as applied to the
calculation of the electromagnetic current and stress-energy correlators
and the extraction of their associated QNM spectra and transport coefficients.
In each case, we first perform the computation in a way that consistently
truncates the result after the linear term in $\gamma$.
Thereafter, we present an alternative calculational scheme
which resums a subset of higher order corrections,
all originating from the $S_{\rm IIB}^{(1)}$ correction term to the
supergravity action.
We emphasize that,
as the explicit forms (and physical effects)
of higher order terms in the supergravity action are presently unknown,
our resummation only captures a limited subset of corrections involving
higher powers of $\gamma$.
Nevertheless, the results of this partial resummation will be seen
to have interesting and suggestive implications for the stability
of holographic predictions at phenomenologically relevant values
of the `t Hooft coupling.

\subsection{Current-current correlator}

We begin by considering correlators of the electromagnetic current operator
$j_{\rm EM}^\mu$ of $\Nfour$ SYM,
defined by gauging a $U(1)$ subgroup of the $SU(4)_{\rm R}$ flavor symmetry.%
\footnote
    {%
    Specifically, we choose the $U(1)$ subgroup for which the
    $\Nfour$ SYM fermions have charges
    $\{ \half, -\half, 0, 0 \}$;
    the explicit form of the current in terms of the $\Nfour$ SYM
    fields is shown in eq.~(2.1) of ref.~\cite{CaronHuot:2006te}.
    }
This current is dual to a $U(1)$ vector field $A_M$ in the
gravitational description.
To compute the two-point correlator
$
    \langle j_{\rm EM}^\mu \, j_{\rm EM}^\nu \rangle
$,
one must solve for the behavior of linearized fluctuations of this
bulk gauge field in the background geometry corresponding to the
equilibrium state of interest.
In the near boundary expansion of the
bulk gauge field $A_M$, the coefficient of the leading term represents
a source coupled to the conserved current $j_{\rm EM}^\mu$, and the
coefficient of the first subleading term encodes the expectation value
of the current in the presence of this source.
Hence, the two-point correlator is given by the variation of the
subleading coefficient with respect to the leading coefficient.

In the $\lambda\to\infty$ limit, the action for the
bulk gauge field is just the standard Maxwell action, which leads
to the usual (curved space) Maxwell equation,
\begin{equation}
    \frac 1{\sqrt{-g}} \,
    \partial_{\mu}(\sqrt{-g} \, g^{\mu\alpha}g^{\nu\beta} \, F_{\alpha\beta})
    = 0 \,.
\label{eq:Maxwell}
\end{equation}
The $\Nfour$ thermal equilibrium state (in the absence of chemical potentials) is dual to the AdS-Schwarzschild geometry, whose metric may be written in the form
\begin{equation}
    ds^2=\frac{r_h^2}{L^2 \, u}
    \left[ -f(u) \, dt^2+dx^2+dy^2+dz^2 \right]
    +\frac{L^2}{4u^2f(u)} \, du^2.
\label{eq:AdS-BH}
\end{equation}
Here, $L$ is the AdS curvature scale, which we choose to set to unity,
while the non-extremality parameter is related to the field theory
temperature $T$ by $r_h \equiv \pi T L^2$.
The coordinate $u$ is an inverted radial coordinate,
with the spacetime boundary lying at $u = 0$;
in terms of a conventional non-inverted radial coordinate~$r$, $u = r_h^2/r^2$.
The blackening function finally reads $f(u) \equiv 1{-}u^2$, and vanishes at the black brane horizon located at $u = 1$.

When evaluated in the above geometry and Fourier transformed with respect to the boundary (Minkowski) coordinates, the Maxwell equation (\ref{eq:Maxwell}) reduces to a pair of
decoupled linear ordinary differential equations for the longitudinal and transverse components of the
electric field~\cite{CaronHuot:2006te},
\begin{align}
    0 &= E_{\bot}''(u)
    +\frac{f'(u)}{f(u)} \, E_{\bot}'(u)+
    \frac{\hat{\omega}^2-\hat{q}^2f(u)}{uf(u)^2} \, E_{\bot} \,,
\label{eq:Etrans}
\\
    0 &= E''_{\|}(u)
    +\frac{\hat{\omega}^2f'(u)}{f(u)(\hat{\omega}^2-\hat{q}^2)} \, E'_{\|}(u)
    +\frac{\hat{\omega}^2-\hat{q}^2f(u)}{uf(u)^2} \, E_{\|} \,,
\label{eq:Elong}
\end{align}
with
$\hat{\omega} \equiv {\omega}/(2 \pi T)$ and
$\hat{q} \equiv {q}/(2 \pi T)$ denoting the rescaled
frequency and spatial wavevector, respectively. Focusing on the transverse electric field
(which determines the transverse part of the current-current correlator
and thus the photoemission spectrum),
the simple field redefinition
$\Psi(u) \equiv \sqrt {f(u)} \, E_\perp(u)$
converts eq.~(\ref{eq:Etrans}) into a
Schrodinger-like equation at zero energy,
\begin{equation}
    -\Psi''(u) + V(u) \, \Psi(u)=0 \,,
\label{psieom}
\end{equation}
with
\begin{equation}
    V(u)
    \equiv
    -\frac{u+\hat{\omega}^2-\hat{q}^2f(u)}{u f(u)^2} \,.
\end{equation}

At non-zero temperature, the retarded current-current correlator 
may be decomposed into transverse and longitudinal pieces via
\begin{equation}
    G^{\rm ret}_{\mu\nu}(\omega,\q)
    =
    P^\perp_{\mu\nu}(\omega,\q) \, \Pi_\perp (\omegahat,\qhat)
    +
    P^\|_{\mu\nu}(\omega,\q) \, \Pi_\| (\omegahat,\qhat) \,,
\end{equation}
with the symmetric projectors defined by
$
    P^\perp_{0\nu}(\omega,\q) \equiv 0
$,
$
    P^\perp_{ij}(\omega,\q) \equiv \delta_{ij} - q_i q_j / q^2
$,
and
$
    P^\|_{\mu\nu}(\omega,\q)
    \equiv
    \eta_{\mu\nu} - Q_\mu Q_\nu / Q^2 - P^\perp_{\mu\nu}(\omega,\q)
$,
where $Q \equiv (\omega,\q)$ and $Q^2 = -\omega^2 + \q^2$.
The transverse correlation function is then given by
\cite{Son:2002}
\begin{equation}
    \Pi_{\bot}(\hat{\omega},\hat{q})
    =
    -\tfrac 18 \, \Nc^2T^2 \lim_{u \to 0}\frac{E_{\bot}'(u)}{E_{\bot}(u)}
    =
    -\tfrac 18 \, {\Nc^2T^2} \lim_{u \to 0}\frac{\Psi'(u)}{\Psi(u)} \,,
\label{eq:Piperp}
\end{equation}
where $E_\perp$ (or $\Psi$) is the solution to eq.~(\ref{eq:Etrans})
[or (\ref{psieom})] satisfying infalling boundary conditions
at the horizon.

Numerous physical observables of interest can be extracted
from the current-current correlator.
Quasinormal modes are poles of the retarded correlator,
regarded as functions of the complex frequency $\omega$ for
fixed wavevector $q$, while pole positions at imaginary values of $q$ and $\omega = 0$
give thermal screening masses \cite{Bak:2007fk}.
The zero-frequency slope, at vanishing wavevector,
determines the electric conductivity,
\begin{equation}
    \sigma \equiv
    - \lim_{\omega \to 0}\mathrm{Im} \,
    \frac{e^2}{\omega} \, \Pi_{\bot}(\omegahat,\qhat{=}0)
    =
    \frac{\Nc^2 \, e^2T}{16 \pi}
    \lim_{\hat{\omega} \to 0}
    \lim_{u \to 0}
	\frac{1}{\hat{\omega}} \,
	\mathrm{Im} \, \frac{\Psi'(u)}{\Psi(u)}
    \,,
\label{eq:sigma}
\end{equation}
where $e$ is the (arbitrarily weak) coupling constant of the electromagnetic
$U(1)$ gauge field coupled to the conserved current.%
\footnote
    {%
    If one regards the $U(1)$ current as a global symmetry current,
    not coupled to a dynamical electromagnetic gauge field,
    then the associated charge diffusion constant is related
    to the conductivity by the Einstein relation,
    $D = \sigma/(e^2 \Xi)$,
    where $\Xi = \frac 18 \, \Nc^2 \, T^2$ is the $\Nfour$ SYM charge
    susceptibility.
    }
Finally, the (equilibrium) photoemission spectrum is determined by the imaginary part
of the transverse correlator evaluated on the lightcone \cite{CaronHuot:2006te},
\begin{equation}
    \frac{d\Gamma_\gamma}{dk}
    =
    \frac{\alpha_{\rm EM}}{\pi} \,
    k \, n_{\rm b}(k) \,
    \left( -4 \, \mathrm{Im} \, \Pi_\perp \right)
    \big|_{\omegahat = \qhat = k/(2\pi T)} \,,
\label{eq:photoemission}
\end{equation}
where $n_{\rm b}(\omega) \equiv (e^{\omega/T} - 1)^{-1}$ is the
usual Bose distribution function.

Expression (\ref{eq:Piperp}) shows that the correlator will have
poles at values of $q$ and $\omega$ for which the denominator,
equal to the boundary value of the electric field (or $\Psi$), vanishes.
In other words, QNMs represent homogeneous
solutions of the bulk Maxwell equations
satisfying infalling boundary conditions at the horizon
and a Dirichlet condition at the spacetime boundary.
At $q = 0$, one may solve the transverse
equation (\ref{eq:Etrans})
and find the resulting roots of $E_\perp$ analytically
\cite{Nunez:2003eq}.
The result is the famous linear spectrum,
\begin{equation}
    \hat{\omega}_n= n \, (\pm 1-i) \,,
\end{equation}
while numerical results for non-zero wavevector may be found in ref.~\cite{Kovtun:2005}.

To incorporate finite coupling corrections in the above calculation,
one begins with the expansion (\ref{eq_1}) of the 10D supergravity action
and retains both the leading and first subleading terms.
Schematically,
\begin{eqnarray}
    S_{\rm IIB}^{(0)}
    &=&
    \frac{1}{2\kappa_{10}}\int d^{10}x \>
    \sqrt{-G}
    \left[
	R_{10}
	-\tfrac 12 {(\partial \phi)^2}
	-\tfrac 1{4\cdot 5!} {(F_5)^2}
    \right],
\\
    S_{\rm IIB}^{(1)}
    &=&
    \frac{L^6}{2\kappa_{10}}\int d^{10}x \>
    \sqrt{-G} \, e^{-\frac 32\phi}
    \left(
	\mathcal{C}+\mathcal{T}
    \right)^4,
\label{eq:SI}
\end{eqnarray}
where
$\kappa_{10}$ is the 10D Newton constant,
$R_{10}$ the 10D Ricci-scalar,
$\phi$  the dilaton field, and
$F_5$ the five-form field strength, while
$\mathcal{C}$ stands for the Weyl tensor,
and $\mathcal{T}$ for a tensor built from the gradient of $F_5$
plus terms quadratic in $F_5$ 
\cite{Gubser:1998nz,
	Paulos:2008tn,
	Myers:2008yi}.
After a reduction to 5D and the extraction of terms
at most quadratic in the emergent 5D bulk gauge field dual
to the EM conserved current, one eventually finds
a $\gamma$-corrected Maxwell equation
which, once again, may be put into the Schrodinger-like form (\ref{psieom})  \cite{Hassanain:2011ce,Vuorinen:2013}.
The needed field redefinition becomes
$
    \Psi(u)=\Sigma(u) E_{\bot}(u)\,,
$
with
\begin{equation}
    \Sigma(u) \equiv \frac{\sqrt{f(u)}}{1+\gamma \, p(u)} \,,\quad
    p(u)=\frac{u^2}{288}
    \left[
	11700
	-343897 \, u^2
	-37760 \, u^3 \, \hat{q}^2
	+87539 \, u^4
    \right] ,
\end{equation}
while the resulting $\gamma$-corrected effective potential reads
\begin{eqnarray}
     V(u)
 &=&
     -\frac{u+\hat{\omega}^2-\hat{q}^2f(u)}{u f(u)^2}
\nonumber \\ &&{}
     +\frac{\gamma}{144 f(u)}
     \Bigl[
	 -11700
	 +2098482 \, u^2
	 -4752055 \, u^4
	+1838319 \, u^6
\nonumber \\ &&\qquad\qquad\quad {}
	+\hat{q}^2
	    \left(
		4770 \, u^{-1}
		+11700 \, u
		-953781 \, u^3+
		1011173 \, u^5
	    \right)
\nonumber \\ &&\qquad\qquad\quad {}
	-\hat{\omega}^2
	    \left(
		4770 \, u^{-1}
		+28170 \, u
		-1199223 \, u^3
	    \right)
    \Bigr] \,.
\label{eq:Vcorrected}
\end{eqnarray}

Given the above potential,
solutions to the equation $\Psi'' = V \, \Psi$ may be expanded in a power series
in $\gamma$.
The black brane horizon at $u = 1$ is a regular singular point of the equation,
where infalling solutions behave locally as $\Psi(u) \sim (1{-}u)^r$ with
characteristic exponent $r = \frac 12 (1-i \omegahat)$.
Hence, one may expand the solutions of interest as
\begin{equation}
    \Psi(u)=(1{-}u)^r\left[\Phi^{(0)}(u) +\gamma \, \Phi^{(1)}(u) +\cdots \right] \,,
\end{equation}
where $\Phi^{(0)}$ and $\Phi^{(1)}$ have the near-horizon
Frobenius expansions,
\begin{equation}
    \Phi^{(0)}(u) \sim \sum_{n=0}^{\infty} \> a_n \, (1{-}u)^n, \hspace{1cm}
    \Phi^{(1)}(u) \sim \sum_{n=0}^{\infty} \> b_n \, (1{-}u)^n \,.
\label{eq:Frob}
\end{equation}
One may further determine the coefficients $\{ a_n \}$ and $\{ b_n \}$ recursively
by inserting these expansions into eq.~(\ref{psieom}) and
collecting like powers of $\gamma$ and $(1{-}u)$.
Without loss of generality, one may set $a_0 = 1$ and $b_0 = 0$.

\begin{figure}
\centering
\hspace*{-5em}
\includegraphics[scale=1.0]{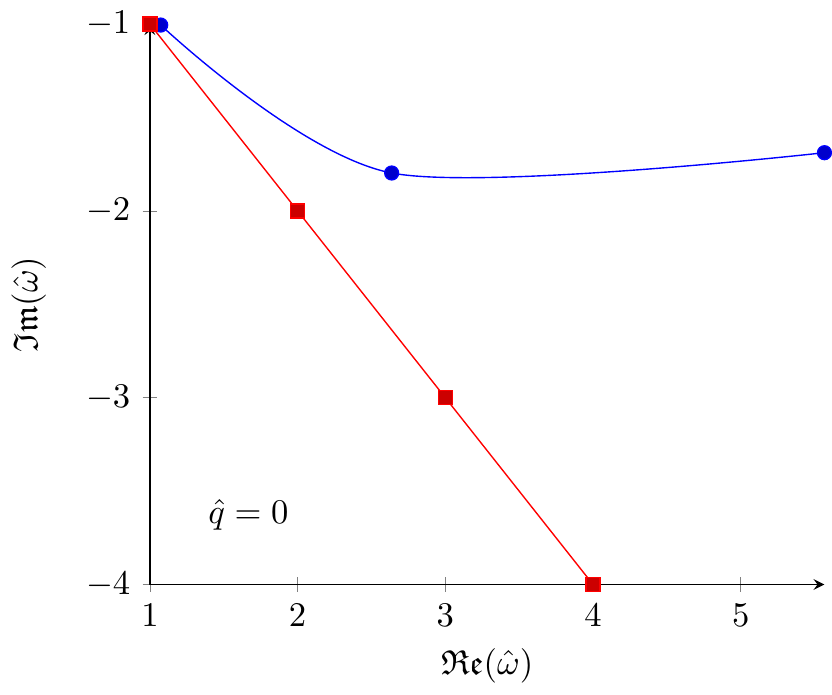}
\includegraphics[scale=1.0]{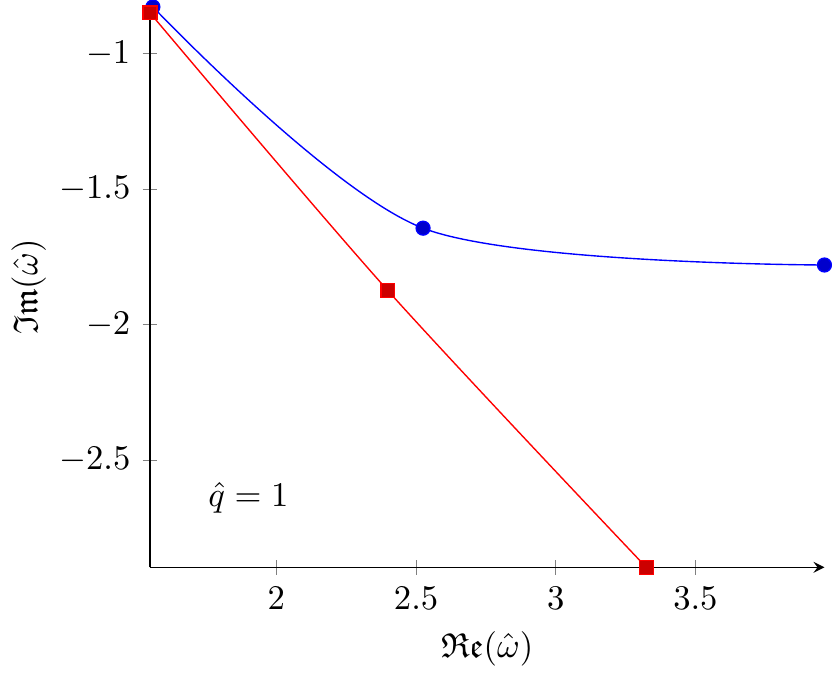}
\hspace*{-4em}
\caption
    {%
    The first few QNM frequencies, divided by $2\pi T$,
    of the electromagnetic current operator
    for $\hat{q}=0$ (left) and $\hat{q}=1$ (right),
    evaluated at $\lambda=\infty$ (red squares) and $\lambda=1000$
    (blue circles).
    The $\lambda  = 1000$ results include the $O(\gamma)$
    corrections, but no higher order contributions.
    Lines have been inserted merely to guide the eye.
    }
\label{fig1}
\end{figure}

The values of frequency, for which the solution also satisfies the Dirichlet condition at the boundary (for a fixed $q$),
may also be expanded in powers of $\gamma$,
\begin{equation}
    \hat{\omega}=\hat{\omega}^{(0)}+\gamma \, \hat{\omega}^{(1)}+\cdots \,.
\end{equation}
If the expansion (\ref{eq:Frob}) is truncated at some upper limit $N$
and used throughout the computational domain $0 \le u \le 1$,
then the Dirichlet condition that $\Psi(0)$ vanish (for all $\gamma$)
reduces to a set of algebraic equations
\begin{equation}
    \sum_{n=0}^{N} \> a_n(\hat{\omega}^{(0)})=0 \,, \hspace{1cm}
    \sum_{n=0}^{N} \> b_n(\hat{\omega}^{(0)},\hat{\omega}^{(1)})=0 \,,
\end{equation}
whose roots yield (approximations to) $\omegahat^{(0)}$ and $\omegahat^{(1)}$.
Carrying out this procedure for values of $N$ sufficiently large that the results
are stable turns out to be a viable computational strategy \cite{Vuorinen:2013}.
Our results for the first few QNMs
are displayed in fig.~\ref{fig1}
and reported in table \ref{table:EM} below,
and fully agree with the findings of ref.~\cite{Vuorinen:2013}.

We have also computed the photoemission spectrum (\ref{eq:photoemission})
by solving the $\gamma$-corrected equation for the
transverse electric field, as in ref.~\cite{Hassanain:2011ce},
and then evaluated the first few moments (\ref{eq:Gamma_n})
of the spectrum,
obtaining the results shown in table \ref{table:1}.

As an alternative to the above approach, in which the QNM frequencies and mode
functions are explicitly expanded in powers of $\gamma$,
one may directly solve the Schrodinger equation (\ref{psieom}) with
the potential (\ref{eq:Vcorrected}) evaluated at some chosen
value of $\gamma$.
Spectral methods provide an efficient numerical approach \cite{Boyd:2001}.
We write
\begin{equation}
    \Psi(u)=(1{-}u)^r \, \Phi(u) \,,
\label{psiphi}
\end{equation}
so that the function $\Phi(u)$ is regular at both the horizon and boundary.
In terms of $\Phi$, the explicit $\gamma$-corrected QNM equation takes the form 
\begin{align}
    -u (1 {+} u) &(1 {-} u^2) \, \Phi''(u)
    + u (1 {+} u)^2 (1 - i\hat{\omega}) \, \Phi'(u)
    + K(u) \, \Phi(u)
    = 0 \,,
\label{eqphi2}
\\ \noalign{\noindent with}
    K(u)
    &\equiv
    (1 - \hat{\omega}^2)
    + \tfrac 14 u (1 - 3 \hat{\omega}^2))
    - \tfrac 14 u^2 (1 + \hat{\omega}^2)
\nonumber \\
     &\quad {}
     + \frac \gamma{144} \, (1 {+} u)
     \Bigl[
	 11700 \, u + 2098482 \, u^3 - 4752055 \, u^5 + 1838319 \, u^7
\nonumber \\
    &\quad\qquad\qquad\qquad {}
    + \qhat^2 \left(4770 + 11700 \, u^2 - 953781 \, u^4 + 1011173 \, u^6 \right)
\nonumber \\
    &\quad\qquad\qquad\qquad {}
	+ \omegahat^2 \left( -4770 - 28170 \, u^2 + 1199223 \, u^4 \right)
    \Bigr] \,.
\label{eq:phieqn}
\end{align}

Next, we expand $\Phi$ in a truncated series of Chebyshev polynomials,
\begin{equation}
    \Phi(u) = \sum_{n=0}^N \> c_n \, T_n(2u{-}1) \,,
\label{eq:Cheb}
\end{equation}
with $T_n(z) \equiv \cos(n \cos^{-1}z)$.%
\footnote
    {%
    The Chebyshev polynomials $\{ T_n(z) \}$ form an orthogonal basis in the
    Hilbert space with inner product
    $
	(f,g) = \int_{-1}^1 dz \> f(z) \, g(z) / \sqrt{1-z^2}
    $.
    }
Requiring equation (\ref{eqphi2}) to be satisfied at the points
$u_j \equiv \frac 12 \, [1-\cos(j\pi/N)]$,
$j =0, 1, \cdots N$,
which comprise a Chebyshev-Gauss-Lobatto grid,
converts the differential equation (\ref{eqphi2}) into a finite
matrix equation of the form $A\cdot v = 0$.
Here, $v$ is a vector
consisting of the $N{+}1$ coefficients $\{ c_n \}$ of the Chebyshev
expansion, while $A$ denotes a matrix whose entries on the $j$'th
row are obtained by evaluating eq.~(\ref{eqphi2}) at the $j$'th grid point.%
\footnote
    {%
    The row of this matrix corresponding to the $u = 0$ endpoint
    automatically enforces the Dirichlet boundary condition $\Phi(0) = 0$.
    }
Nonvanishing solutions of this homogeneous set of equations only exist when
\begin{equation}
    \det(A)=0 \,.
\label{det}
\end{equation}
This determinant is a polynomial in $\omegahat$, whose roots
$\{ \omegahat_k^{(N)} \}$
rapidly converge (for fixed $k$) as the number of grid points $N$ increases.
Evaluating these roots numerically is straightforward
(for relatively modest values of $N$)
given specific values of the wavevector $\qhat$
and the parameter $\gamma$.

\begin{figure}
\centering
\hspace*{-5em}
\includegraphics[scale=1.0]{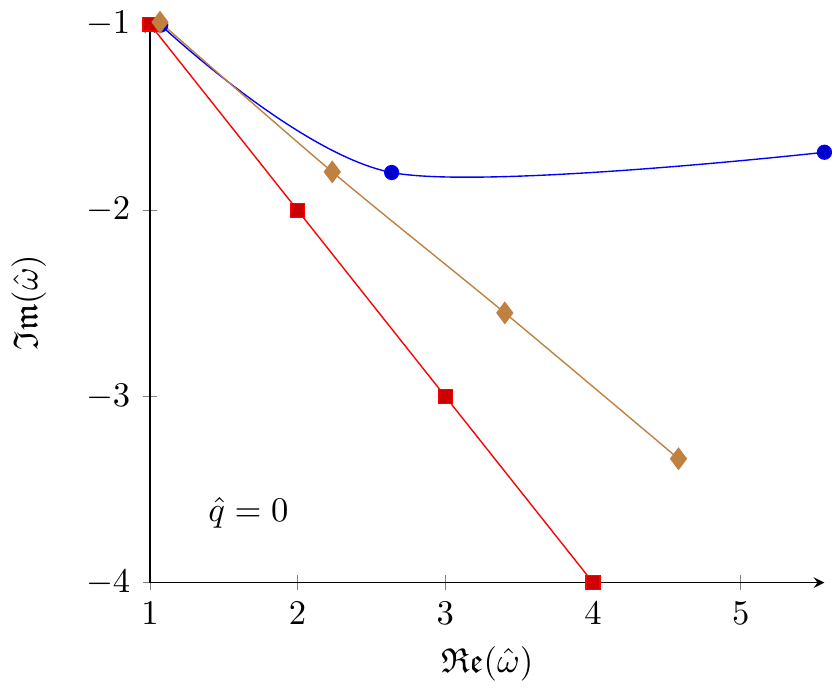}
\includegraphics[scale=1.0]{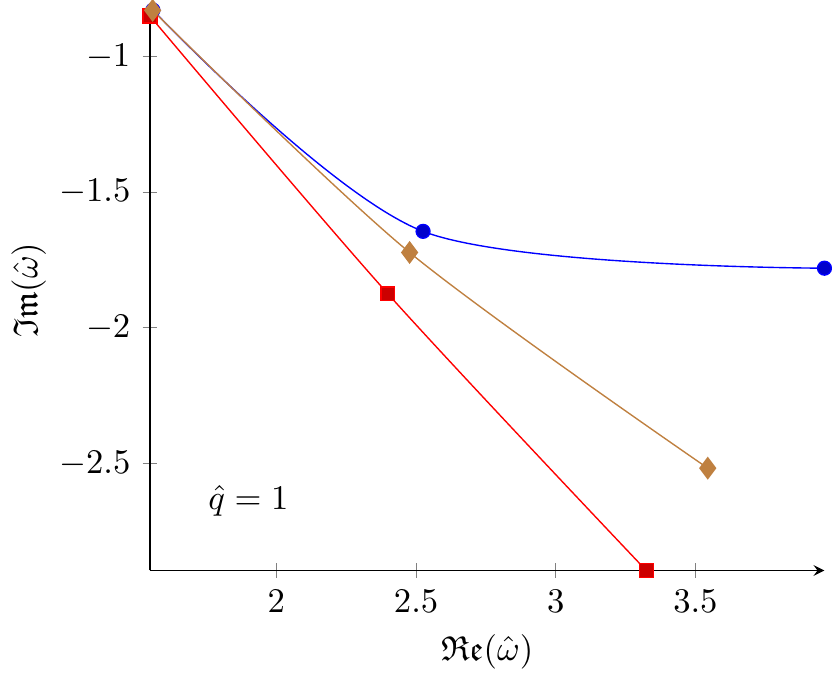}
\hspace*{-5em}
\caption
    {%
    The first few QNM frequencies, divided by $2\pi T$,
    of the electromagnetic current operator
    for $\hat{q}=0$ (left figure) or $\hat{q}=1$ (right figure)
    at $\lambda = 1000$.
    Results obtained by directly solving the QNM equation at this value
    of $\lambda$
    using spectral methods are shown as brown diamonds,
    while the red squares and blue circles show the same zeroth and first order
    results, respectively,
    previously displayed in fig.~\ref{fig1}.
    Again, lines merely serve to guide the eye.
    }\label{fig2}
\end{figure}

\begin{table}
\begin{center}

\hspace*{-2em}
\begin{tabular}{|c||c|c|c||c|c|c|}
\hline
& \multicolumn{3}{|c||}{$\omegahat_k^{\rm EM}(\qhat = 0)$} 
& \multicolumn{3}{c|}{$\omegahat_k^{\rm EM}(\qhat = 1)$} 
\\\hline
$k$ & $O(\gamma^0)$ & $O(\gamma^1)$ & resummed &
      $O(\gamma^0)$ & $O(\gamma^1)$ & resummed 
\\\hline
1 & $1{-}i$ & $1.073 - 1.005i$ & $1.068 - 0.990i$ & $1.547 - 0.850i$ & $1.558 - 0.828i$ & $1.557 - 0.829i$
\\
2 & $2{-}2i$ & $2.637 - 1.797i$ & $2.237 - 1.794i$ & $2.399 - 1.874i$ & $2.525 - 1.645i$ & $2.477 - 1.722i$
\\
3 & $3{-}3i$ & $5.536 - 1.692i$ & $3.403 - 2.551i$ & $3.323 - 2.895i$ & $3.957 - 1.791i$ & $3.544 - 2.518i$
\\
4 & $4{-}4i$ & $11.07  + 0.47i$ & $4.57 - 3.34 i$ & $4.28 - 3.91i$ & $6.37 - 0.39i$ & $4.67 - 3.31i$
\\\hline
\end{tabular}
\hspace*{-2em}
\end{center}
\caption
    {%
    The first four QNM frequencies, divided by $2\pi T$,
    of the electromagnetic current operator
    $\{ \omegahat_k^{\rm EM} \}$, $k = 1, \cdots 4$,
    for $\qhat = 0$ (left) and $\qhat = 1$ (right),
    at $\lambda = 1000$.
    Respective columns show the results from the
    zeroth order, first order, and resummed approximations
    discussed in the text.
    \label{table:EM}
    }
\end{table}

\begin{table}
\begin{center}

\begin{tabular}{|c||c|c||c|c|}
\hline
& \multicolumn{2}{|c||}{$\lambda_{\text{breakdown}}(\qhat=0)$} 
& \multicolumn{2}{c|}{$\lambda_{\text{breakdown}}(\qhat=1)$} 
\\\hline
$k$ & $O(\gamma^1)$ & resummed  & $O(\gamma^1)$ & resummed 
\\\hline
 1&139& $\approx 22.5$ & 57.1 & $< 5$
\\
 2&382.2 & $\approx 21.7$& 195& $\approx 8$
\\
 3&767.5 & $\approx 21.3$  & 437 & $\approx 14$
\\
 4&1298 & $\approx 21.1$ & 793 & $\approx 16$
\\\hline
\end{tabular}
\end{center}
\caption
    {%
    Values of $\lambda$ below which the deviation of the 
    QNM frequency $\omega_k^{\rm EM}$
    from its $\lambda = \infty$ limit exceeds the
    $\lambda = \infty$ value.
    Respective columns show the results obtained using either the
    first order or resummed approximations for the QNM frequency.
    \label{table:breakdown}
    }
\end{table}

Solving the QNM equation in the fashion described above yields values for the
quasinormal mode frequencies with nonlinear dependence on $\gamma$.
One has, in effect, resummed a subset of higher order contributions to the QNM
frequencies which arise solely from the first order,
i.e.~${\mathcal O}(\gamma)$, correction to the supergravity action.
One may hope --- although there is no guarantee --- that this is the
dominant source of all higher order contributions.

In figure \ref{fig2} and  table \ref{table:EM},
we display the effects of the resummation
for the first few QNM frequencies, again
at $\lambda = 1000$ and both $\qhat = 0$ and $\qhat = 1$,
and compare the results to the previous unresummed values.
As can be readily verified, the size of the $O(\gamma)$ correction increases rapidly with
the QNM mode number $k$, asymptotically growing like $k^4$.
This reflects the fact that finite coupling corrections arise from
higher dimension operators in the supergravity action (\ref{eq:SI}).
Although the size of the $O(\gamma)$ correction to the first QNM frequency
appears modest at $\lambda = 1000$,
as seen in the top row of table \ref{table:EM},
the $O(\gamma)$ correction exceeds the leading $\lambda=\infty$ value 
of the first QNM frequency at $\lambda = 139$ (for $q=0$),
or $\lambda = 57.1$ (for $q=1$),
above our benchmark phenomenological range.
For the second mode (as noted previously) and all higher modes,
the ``breakdown'' values of $\lambda$,
below which the $O(\gamma)$ correction exceeds the leading term,
are even larger.

In contrast, our resummed approximation for the QNM frequencies
yields results which deviate from the $\lambda = \infty$ values
substantially less.
As a concrete measure of this, table~\ref{table:breakdown} compares, for the
first few modes, the breakdown values of $\lambda$ below which
the deviation of the QNM frequency from its $\lambda = \infty$ value
exceeds the $\lambda = \infty$ result, in either scheme.%
\footnote
    {%
    When computing resummed approximations,
    the convergence of both the Frobenius (\ref{eq:Frob}) and spectral
    (\ref{eq:Cheb}) expansions progressively degrade as the value of
    $\gamma$ is increased,
    making precise determinations of breakdown values of $\lambda$
    challenging.
    }
For the resummed approximation, we find that these nominal breakdown
values of $\lambda$ are substantially smaller than the breakdown values
of the first order results.
For the first four modes, the breakdown values of the resummed
approximation lie within or below our benchmark phenomenological
range of 10--40.
Moreover, the breakdown
values of the resummed approximation grow far less rapidly with
mode number than do the $O(\gamma)$ results.

One may also apply our partial resummation scheme when evaluating the
zero-frequency slope of the correlator which determines
the electric conductivity of the ${\mathcal N}=4$ SYM plasma
via eq.~(\ref{eq:sigma}).
However, since the conductivity receives a much smaller
$O(\gamma)$ correction than do the QNM frequencies,
the effect of the resummation is much more modest for the conductivity.
At, for example, $\lambda = 1000$ the relative size of the $O(\gamma)$
correction to the conductivity is about $8 \times 10^{-3}$, and our
resummation increases this deviation from the $\lambda = \infty$ value
by a further $5 \times 10^{-5}$.
In contrast to the situation with QNM frequencies,
the nominal breakdown value of the resummed approximation
for the conductivity is somewhat larger than the breakdown value
of the first order approximation
($\lambda \approx 57.9$ instead of 39.7).

\subsection{Stress-energy correlator \label{scalar}}

The dynamics of
linearized metric perturbations determine the
stress-energy tensor correlator.
We will focus on the $\ell = 2$ or shear channel, for which it is
sufficient to consider $\delta g_{xy}$ as the only non-zero component
of the perturbation when the wavevector points in the $z$-direction.%
\footnote
    {%
    Note that in ref.~\cite{Kovtun:2005},
    the channel with $\ell = 2$ rotational symmetry about the wavevector
    was referred to as the ``scalar channel''
    because the corresponding metric perturbation
    satisfies the same equation as a minimally coupled massless scalar,
    and the $\ell = 1$ or vector channel was referred to as
    the ``shear channel''.
    Our terminology is motivated by the fact that the shear viscosity
    can be obtained from the zero-frequency limit of the correlator
    in the $\ell = 2$ channel. 
    }
After Fourier transforming with respect to the boundary coordinates,
the rescaled perturbation
\begin{equation}
    Z(u) \equiv \frac{u}{r_h^2} \, \delta g_{xy}(u) 
\end{equation}
satisfies the ${\mathcal O}(\gamma)$ corrected equation of motion
\cite{Benincasa},
\begin{equation}
     -Z''(u)
     + P(u) \, Z'(u)
     + Q(u) \, Z(u) = 0 \,,
\label{eq:Z1eq}
\end{equation}
with
\begin{align}
    P(u)
    &\equiv \frac{1+u^2}{u f(u)}+
    \tfrac 14 \, \gamma
    \left(
	  600 \,u
	- 2306 \, u^3
	- 3171 \, u^5
	- 3840 \, \qhat^2 \, u^4
    \right)
\label{eq:P}
\\
\noalign{\noindent and}
    Q(u)
    &\equiv
    - (1+ 30 \, \gamma) \,
	\frac{\omegahat^2 - \qhat^2 f(u)}{u f(u)^2}
\nonumber\\ & \quad {}
    -  \frac{\gamma}{4 f(u)^2} \,
    \Bigl[
	    50 \, u^2
	    - 275 \, u^6
	    + 225 \, u^8
\nonumber\\ & \qquad\qquad\quad {}
	+ \omegahat^2
	    \left(
	    600 \, u 
	    - 2856 \, u^3
	    + 2136 \, u^5
	    \right)
\nonumber\\[3pt] & \qquad\qquad\quad {}
	+ \qhat^2
	    \left(
	    - 300 \, u
	    + 3456 \, u^3
	    - 6560 \, u^5
	    + 3404 \, u^7
	    \right)
\nonumber\\[3pt] & \qquad\qquad\quad {}
	+ \qhat^4
	    \left( 768 \, u^4 + 768 \, u^6 \right)
    \Bigr] \,.
\label{eq:Q}
\end{align}

\begin{figure}
\centering
\includegraphics[scale=1.0]{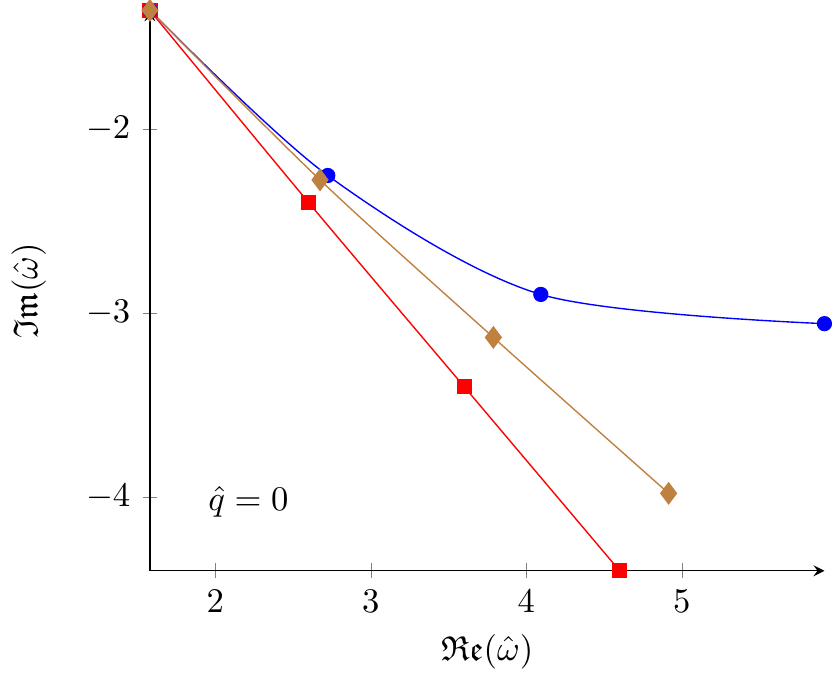}
\caption
    {%
    The first few QNM frequencies, divided by $2\pi T$,
    for the shear channel of the stress-energy correlator,
    evaluated for $\hat{q}=0$ and $\lambda = 500$.
    Results obtained by directly solving the QNM equation at this value
    of $\lambda$
    using spectral methods are shown as brown diamonds,
    while the red squares and blue circles show results truncated
    at zeroth and first order in $\gamma$, respectively.
    As before, lines merely serve to guide the eye.
    }
\label{fig3}
\end{figure}

\begin{table}
\begin{center}

\begin{tabular}{|c||c|c|c|}
\hline
& \multicolumn{3}{|c|}{$\omegahat_k^{\rm shear}(\qhat = 0)$} 
\\\hline
$k$ & $O(\gamma^0)$ & $O(\gamma^1)$ & resummed
\\\hline
1 & $1.560 - 1.373i$ & $1.581-1.356i$  & $1.579-1.356i$ 
\\
2 & $2.585 - 2.382i$ & $2.723-2.253i$& $2.673-2.277i$ 
\\
3 & $3.594 - 3.385i$ & $4.093-2.899i$    & $3.789-3.133i$
\\
4 & $4.60 - 4.39i$ & $5.92-3.07i$ & $4.91-3.98i$ 
\\\hline
\end{tabular}
\end{center}
\caption
    {%
    The first four QNM frequencies, divided by $2\pi T$,
    $\{ \hat{\omega}_k^{\rm shear} \}$, $k = 1, \cdots 4$,
    in the shear channel of the stress-energy correlator,
    for $\qhat = 0$ and $\lambda = 500$.
    Respective columns show the results from the
    zeroth order, first order, and resummed approximations
    discussed in the text.
    \label{table:shear}
    }
\end{table}

\begin{table}
\begin{center}

\begin{tabular}{|c||c|c|}
\hline
& \multicolumn{2}{|c|}{$\lambda_{\rm breakdown}$}
\\\hline
$k$ & $O(\gamma^1)$ & resummed 
\\\hline
1 & 27.6 & $<1$
\\
2 & 71.2 & $<1$ 
\\
3 & 135.5 & $<1$ 
\\
4 & 221 & $<1$
\\\hline
\end{tabular}
\end{center}
\caption
    {%
    Values of $\lambda$ below which the deviation of the 
    QNM frequency $\omega_k^{\rm shear}$
    from its $\lambda = \infty$ limit exceeds the
    $\lambda = \infty$ value.
    Respective columns show the results obtained using either the
    first order or resummed approximations for the QNM frequency.
    \label{table:breakdown2}
    }
\end{table}

Boundary conditions for finding quasinormal modes are the same
as discussed earlier: infalling behavior at the horizon and Dirichlet
at the boundary. Solutions up to $O(\gamma)$ of the above equation were found in ref.~\cite{Stricker:2013lma} using the
Frobenius expansion technique discussed in the previous subsection.
One may, however, also solve the equation directly for specific
values of $\gamma$, just as we did for the electromagnetic current correlator,
using spectral methods.
Figure~\ref{fig3} shows a comparison of
the results of the two approaches for $\lambda = 500$
and $\qhat = 0$, with table \ref{table:shear} listing explicit values.
Once again, at this value of $\lambda$
(which is still well above the phenomenologically relevant range)
we observe a substantial difference between the $O(\gamma)$ results
and our resummed values, with the resummation decreasing the difference
from the $\lambda = \infty$ limit.
In parallel with the previous subsection,
we show in table~\ref{table:breakdown2}
the breakdown values of $\lambda$,
computed with both first order and resummed approximations,
below which the first few shear channel QNM frequencies deviate
from their $\lambda = \infty$ limits by more than
their $\lambda = \infty$ values.
We find that for the shear channel quasinormal frequencies,
our resummation scheme leads to nominal breakdown values of $\lambda$
for all modes up to $k=4$ lying below our
benchmark phenomenological range of 10--40.

Similarly to the plasma conductivity considered earlier,
the shear viscosity is determined by the zero-frequency slope of the 
retarded correlator of $T_{xy}$
evaluated at vanishing wavenumber,
\begin{equation}
    \eta = -\lim_{\omega \to 0} \> \mathrm {Im}
    \frac{1}{\omega} \, G_{xy,xy}^{\rm ret}(\omega,0) \,.
\label{eq:eta}
\end{equation}
Including $O(\gamma)$ corrections, 
it can be shown that this correlator is given by
\cite{Buchel:2005}
\begin{equation}
    G_{xy,xy}^{\rm ret}(\omega,q)
    =
    \lim_{u\to 0} \frac{\Nc^2 \, (r^0_h)^4}{4\pi^2} \, \frac{Z'(u)}{u \, Z(u)},
\label{eq:Gxyxy}
\end{equation}
where $Z(u)$ is a solution to eqs.~(\ref{eq:Z1eq})--(\ref{eq:Q})
at $\qhat=0$ that satisfies infalling boundary conditions at the horizon,
and
$
    r^0_h = \pi T L^2 / (1 + 15 \, \gamma)
$
is the horizon position of the $\gamma = 0$ geometry.%
\footnote
    {%
    This $\gamma = 0$ horizon position,
    used consistently in ref.~\cite{Buchel:2005},
    differs from the horizon position
    $
	r_h = \pi T L^2 / (1 + \frac{265}{16} \, \gamma)
    $
    of the $\gamma$-corrected metric
    derived in ref.~\cite{Pawelczyk:1998pb} and used throughout
    refs.~\cite{Hassanain:2012uj,Hassanain:2011fn,Hassanain:2011ce}.
    }
Using the Frobenius method to solve for the metric perturbation to linear order in $\gamma$ now leads to
\begin{equation}
    G_{xy,xy}^{\rm ret}(\omega,0)
    =
    -i\frac{\Nc^2 \, (r^0_h)^4 \, \omega}{8\pi^3 T} \,
    (1+195 \, \gamma)+\mathcal{O}(\omega^2,\gamma^2) \,,
\end{equation}
from which one can extract the result found in ref.~\cite{Buchel:2008sh},
$
    \eta
    = \tfrac{\pi}{8} \, \Nc^2 \, T^3 \, ( 1+135  \, \gamma+ \cdots )
$.

One may also apply our partial resummation scheme when evaluating the
zero-frequency slope of the correlator (\ref{eq:Gxyxy}).
As with the conductivity,
since the shear viscosity receives a much smaller
$O(\gamma)$ correction than do the QNM frequencies,
the effect of the resummation is much more modest for the this
transport coefficient.
At, for example, $\lambda = 1000$ the relative size of the $O(\gamma)$
correction to the viscosity is about $6 \times 10^{-4}$, and our
resummation increases this deviation from the $\lambda = \infty$ value
by a mere $10^{-7}$.
As with conductivity,
the resummed approximation for the shear viscosity
leads to a somewhat larger nominal breakdown value of $\lambda$
as compared to the first order approximation
($\lambda \approx 14$ instead of $\approx 7$).

\section {Conclusions}
\label{sec:conclusion}

Our examination of finite-$\lambda$ corrections to holographic results
for thermal quantities is motivated by an obvious desire to understand
more clearly the applicability of gauge/gravity duality to the physics
of quark-gluon plasma as produced in real heavy ion collisions.
In particular,
how large an error is made when the system is modeled as infinitely
strongly coupled, as is customary in most holographic calculations
of non-equilibrium dynamics performed in the supergravity limit?
We approached this problem by comparing the behaviors of
the strong coupling expansions for a variety of thermal quantities,
for which the leading order finite coupling correction,
of order $\lambda^{-3/2}$ in the 't Hooft coupling, is known.
Our comparison has singled out quasinormal mode frequencies as quantities
for which the finite coupling corrections appear particularly problematic at
phenomenologically interesting values of the 't Hooft coupling.
We discovered, however, that a partial inclusion of higher order contributions,
generated by the leading corrections to the supergravity action,
leads to a dramatic reduction in the predicted size of finite coupling
effects in quasinormal mode frequencies.

One clear, but unsurprising, message of our comparison is that
notions of strong (or weak) coupling domains can depend rather
strongly on the physics observable of interest.
For some quantities, 't Hooft couplings of order 10--40 may be easily
accessible via a one or two term strong coupling expansion,
while for other quantities this is clearly not the case.
The full reasons underlying such behavior are unclear,
but some insight may perhaps be drawn from similar issues at weak coupling.
There, an extensive amount of work has been devoted to the calculation of
high order perturbative results for a number of equilibrium
thermodynamic quantities.
An important lesson from this work is that the convergence of
perturbation theory is intimately related to the relative influence
of different momentum or energy scales contributing to the observable
in question.
Weak coupling expansions for physical quantities that are
dominantly sensitive to the hard thermal scale of $2\pi T$ are found
to behave significantly better than expansions of quantities
which are more sensitive to the soft $gT$ and ultrasoft $g^2 T$ scales 
which originate from electro- and magnetostatic screening, respectively.
The degraded perturbative stability arises from the dependence on
the ratio of these scales, which appears in the form of
contributions suppressed only by single powers (and logarithms) of $g$,
instead of $\alpha_s = g^2/(4\pi)$.
Ultimately, this reflects the diverging infrared sensitivity of the
Bose-Einstein distribution of gluonic fields.

Whether it is possible to interpret differences in stability of
strong coupling expansions in an analogous manner remains to be seen.
However,
considering the fact that the ${\mathcal O}(\alpha'^3)$ corrections to
$\lambda=\infty$ results originate from higher derivative operators added
to the supergravity action, and that the leading finite coupling corrections to
the QNM frequencies can be seen to grow rapidly with the mode number,
perhaps such expectations are not completely unfounded.
In the meantime,
trying to generalize many more holographic calculations 
to include at least the leading finite 't Hooft couplings 
would clearly be worthwhile.

\begin{acknowledgments}
The authors would like to thank Andreas Karch, Andrei Starinets, and
Stefan Stricker for useful discussions. 
The work of A.V.~was supported by the Academy of Finland, grant Nr. 273545.
The work of L.Y.~was supported, in part, by the U.S. Department
of Energy under Grant No.~DE-SC0011637, and by
the Alexander von Humboldt Foundation.
L.Y.~thanks the University of Regensburg and the Alexander von Humboldt
Foundation for generous support and hospitality during portions of
this work.
A.V. and S.W.~thank the Institute for Nuclear Theory at the University of
Washington for its hospitality and support during the completion of this work.

\end{acknowledgments}


\begin{thebibliography}{99}

\bibitem{Shuryak:2004cy}
  E.~V.~Shuryak,
  {\it What RHIC experiments and theory tell us about properties of quark-gluon plasma?,}
  \npa {750}{2005}{64},
  \hepph{0405066}.


\bibitem{Brambilla:2014jmp}
  N.~Brambilla, S.~Eidelman, P.~Foka, S.~Gardner, A.~S.~Kronfeld, M.~G.~Alford, R.~Alkofer and M.~Butenschoen {\it et al.},
  {\it QCD and strongly coupled gauge theories: challenges and perspectives,}
  \epjc{74}{2014}{2981},
  \arXivid{1404.3723} [hep-ph].

\bibitem{Gale:2013da}
  C.~Gale, S.~Jeon and B.~Schenke,
  {\it Hydrodynamic modeling of heavy-ion collisions,}
  \ijmpa{28}{2013}{1340011},
  \arXivid{1301.5893} [nucl-th].

\bibitem{Kovtun:2004de} 
  P.~Kovtun, D.~T.~Son and A.~O.~Starinets,
  {\it Viscosity in strongly interacting quantum field theories
  from black hole physics,}
  \prl {94}{2005}{111601},
  \hepth{0405231}.

\bibitem{Laermann:2003cv}
  E.~Laermann and O.~Philipsen,
  {\it The status of lattice QCD at finite temperature,}
  \arnps{53}{2003}{163},
  \hepph{0303042}.


\bibitem{Gubser:2009md}
  S.~S.~Gubser and A.~Karch,
  {\it From gauge-string duality to strong interactions: a pedestrian's guide,}
  \arnps {59}{2009}{145},
  \arXivid{0901.0935} [hep-th].

\bibitem{CasalderreySolana:2011us}
  J.~Casalderrey-Solana, H.~Liu, D.~Mateos, K.~Rajagopal and U.~A.~Wiedemann,
  {\it Gauge/string duality, hot QCD and heavy ion collisions,}
  \arXivid{1101.0618} [hep-th].

\bibitem{Chesler:2015lsa}
  P.~M.~Chesler and W.~van der Schee,
  {\it Early thermalization, hydrodynamics and energy loss in AdS/CFT,}
  \arXivid{1501.04952} [nucl-th].

\bibitem{Herzog:2006gh}
  C.~P.~Herzog, A.~Karch, P.~Kovtun, C.~Kozcaz and L.~G.~Yaffe,
  {\it Energy loss of a heavy quark moving through $\Nfour$ supersymmetric Yang-Mills plasma,}
  \jhep {0607}{2006}{013},
  \hepth{0605158}.

\bibitem{Caceres:2006as}
  E.~Caceres and A.~Guijosa,
  {\it On drag forces and jet quenching in strongly coupled plasmas,}
  \jhep {0612}{2006}{068},
  \hepth{0606134}.

\bibitem{Caceres:2006dj}
  E.~Caceres and A.~Guijosa,
  {\it Drag force in charged $\Nfour$ SYM plasma,}
  \jhep {0611}{2006}{077},
  \hepth{0605235}.

\bibitem{Gubser:2006bz}
  S.~S.~Gubser,
  {\it Drag force in AdS/CFT,}
  \prd {74}{2006}{126005},
  \hepth{0605182}.

\bibitem{CasalderreySolana:2007qw}
  J.~Casalderrey-Solana and D.~Teaney,
  {\it Transverse momentum broadening of a fast quark in a $\Nfour$ Yang
    Mills Plasma,}
  \jhep {0704}{2007}{039},
  \hepth{0701123}.


\bibitem{Chesler:2008uy}
  P.~M.~Chesler, K.~Jensen, A.~Karch and L.~G.~Yaffe,
  {\it Light quark energy loss in strongly-coupled $\Nfour$ supersymmetric Yang-Mills plasma,}
  \prd {79}{2009}{125015},
  \arXivid{0810.1985} [hep-th].

\bibitem{Gubser:2008as}
  S.~S.~Gubser, D.~R.~Gulotta, S.~S.~Pufu and F.~D.~Rocha,
  {\it Gluon energy loss in the gauge-string duality,}
  \jhep {0810}{2008}{052},
  \arXivid{0803.1470} [hep-th].

\bibitem{Chesler:2008wd}
  P.~M.~Chesler, K.~Jensen and A.~Karch,
  {\it Jets in strongly-coupled $\Nfour$ super Yang-Mills theory,}
  \prd {79}{2009}{025021},
  \arXivid{0804.3110} [hep-th].

\bibitem{Chesler:2014jva}
  P.~M.~Chesler and K.~Rajagopal,
  {\it Jet quenching in strongly coupled plasma,}
  \prd {90}{2014}{025033},
  \arXivid{1402.6756} [hep-th].

\bibitem{CaronHuot:2006te}
  S.~Caron-Huot, P.~Kovtun, G.~D.~Moore, A.~Starinets and L.~G.~Yaffe,
  {\it Photon and dilepton production in supersymmetric Yang-Mills plasma,}
  \jhep {0612}{2006}{015},
  \hepth{0607237}.

\bibitem{Hassanain:2012uj}
  B.~Hassanain and M.~Schvellinger,
  {\it Plasma photoemission from string theory,}
  \jhep {1212}{2012}{095},
  \arXivid{1209.0427} [hep-th].

\bibitem{Baier:2012tc}
  R.~Baier, S.~A.~Stricker, O.~Taanila and A.~Vuorinen,
  {\it Holographic dilepton production in a thermalizing plasma,}
  \jhep {1207}{2012}{094},
  \arXivid{1205.2998} [hep-ph].

\bibitem{Baier:2012ax}
  R.~Baier, S.~A.~Stricker, O.~Taanila and A.~Vuorinen,
  {\it Production of prompt photons: holographic duality and thermalization,}
  \prd {86}{2012}{081901},
  \arXivid{1207.1116} [hep-ph].

\bibitem{Muller:2012rh}
  B.~Muller and D.~L.~Yang,
  {\it Production of prompt photons and dileptons in rapid holographic thermalization,}
  \arXivid{1212.3354}.

\bibitem{Chesler:2008hg}
  P.~M.~Chesler and L.~G.~Yaffe,
  {\it Horizon formation and far-from-equilibrium isotropization in supersymmetric Yang-Mills plasma,}
  \prl {102}{2009}{211601},
  \arXivid{0812.2053} [hep-th].

\bibitem{Heller:2012km}
  M.~P.~Heller, D.~Mateos, W.~van der Schee and D.~Trancanelli,
  {\it Strong coupling isotropization of non-Abelian plasmas simplified,}
  \prl {108}{2012}{191601},
  \arXivid{1202.0981} [hep-th].

\bibitem{Heller:2013oxa}
  M.~P.~Heller, D.~Mateos, W.~van der Schee and M.~Triana,
  {\it Holographic isotropization linearized,}
  \jhep {1309}{2013}{026},
  \arXivid{1304.5172} [hep-th].

\bibitem{Fuini:2015hba}
  J.~F.~Fuini and L.~G.~Yaffe,
  {\it Far-from-equilibrium dynamics of a strongly coupled non-Abelian plasma with non-zero charge density or external magnetic field,}
  \jhep {1507}{2015}{116},
  \arXivid{1503.07148} [hep-th].

\bibitem{Chesler:2009cy}
  P.~M.~Chesler and L.~G.~Yaffe,
  {\it Boost invariant flow, black hole formation, and far-from-equilibrium
  dynamics in $\Nfour$ supersymmetric Yang-Mills theory,}
  \prd {82}{2010}{026006},
  \arXivid{0906.4426} [hep-th].

\bibitem{Heller:2011ju}
  M.~P.~Heller, R.~A.~Janik and P.~Witaszczyk,
  {\it The characteristics of thermalization of boost-invariant plasma
  from holography,}
  \prl {108}{2012}{201602},
  \arXivid{1103.3452} [hep-th].

\bibitem{Bantilan:2012vu}
  H.~Bantilan, F.~Pretorius and S.~S.~Gubser,
  {\it Simulation of asymptotically AdS$_5$ spacetimes with a
  generalized harmonic evolution scheme,}
  \prd {85}{2012}{084038},
  \arXivid{1201.2132} [hep-th].



\bibitem{Bantilan:2014sra}
  H.~Bantilan and P.~Romatschke,
  {\it Simulation of black hole collisions in asymptotically anti de
  Sitter spacetimes,}
  \prl {114}{2015}{081601},
  \arXivid{1410.4799} [hep-th].

\bibitem{Wu:2013qi}
  B.~Wu,
  {\it On holographic thermalization and gravitational collapse of tachyonic scalar fields,}
  \jhep {1304}{2013}{044},
  \arXivid{1301.3796} [hep-th].

\bibitem{Danielsson:1999zt}
  U.~H.~Danielsson, E.~Keski-Vakkuri and M.~Kruczenski,
  {\it Spherically collapsing matter in AdS, holography, and shellons,}
  \npb {563}{1999}{279},
  \hepth{9905227}.

\bibitem{Taanila:2015sda}
  O.~Taanila,
  {\it Holographic thermalization and Oppenheimer-Snyder collapse,}
  \arXivid{1507.00878} [hep-th].

\bibitem{Chesler:2010bi}
  P.~M.~Chesler and L.~G.~Yaffe,
  {\it Holography and colliding gravitational shock waves in asymptotically AdS$_5$ spacetime,}
  \prl {106}{2011}{021601},
  \arXivid{1011.3562} [hep-th].

\bibitem{Casalderrey-Solana:2013aba}
  J.~Casalderrey-Solana, M.~P.~Heller, D.~Mateos and W.~van der Schee,
  {\it From full stopping to transparency in a holographic model of heavy ion collisions,}
  \prl {111}{2013}{181601},
  \arXivid{1305.4919} [hep-th].

\bibitem{Casalderrey-Solana:2013sxa}
  J.~Casalderrey-Solana, M.~P.~Heller, D.~Mateos and W.~van der Schee,
  {\it Longitudinal coherence in a holographic model of asymmetric collisions,}
  \prl {112}{2014}{221602},
  \arXivid{1312.2956} [hep-th].

\bibitem{Chesler:2015fpa}
  P.~M.~Chesler, N.~Kilbertus and W.~van der Schee,
  {\it Universal hydrodynamic flow in holographic planar shock collisions,}
  \arXivid{1507.02548} [hep-th].

\bibitem{Chesler:2015wra}
  P.~M.~Chesler and L.~G.~Yaffe,
  {\it Holography and off-center collisions of localized shock waves,}
  \arXivid{1501.04644} [hep-th].

\bibitem{Chesler:2015bba}
  P.~M.~Chesler,
  {\it Colliding shockwaves and hydrodynamics in extreme conditions,}
  \arXivid{1506.02209} [hep-th].

\bibitem{Panero:2009tv}
    M.~Panero,
    {\it Thermodynamics of the QCD plasma and the large-$N$ limit,}
    \prl {103}{2009}{232001},
    \arXivid{0907.3719}.

\bibitem{Bali:2013kia}
    G.~S.~Bali, F.~Bursa, L.~Castagnini, S.~Collins, L.~Del Debbio,
    B.~Lucini and M.~Panero,
    {\it Mesons in large-N QCD,}
    \jhep {1306}{2013}{071},
    \arXivid{1304.4437}.

\bibitem{Gubser:1998nz}
    S.~S.~Gubser, I.~R.~Klebanov and A.~A.~Tseytlin,
    {\it Coupling constant dependence in the thermodynamics of
    $\Nfour$ supersymmetric Yang-Mills theory,}
    \npb {534}{1998}{202},
    \hepth{9805156}.


\bibitem{Buchel:2005}
    A.~Buchel, J.~T.~Liu, A.~O.~Starinets,
    {\it Coupling constant dependence of the shear viscosity in
    $\Nfour$ supersymmetric Yang-Mills theory,}
    \npb {707}{2005}{56-68},
    \hepth{0406264}.

\bibitem{Buchel:2008sh} 
  A.~Buchel,
  {\it Resolving disagreement for $\eta/s$ in a CFT plasma at finite coupling,}
  \npb {803}{2008}{166},
  \arXivid{0805.2683} [hep-th].

\bibitem{Hassanain:2011fn}
    B.~Hassanain and M.~Schvellinger,
    {\it Plasma conductivity at finite coupling,}
    \jhep {1201}{2012}{114},
    \arXivid{1108.6306}.

\bibitem{Hassanain:2011ce}
    B.~Hassanain and M.~Schvellinger,
    {\it Diagnostics of plasma photoemission at strong coupling,}
    \prd {85}{2012}{086007},
    \arXivid{1110.0526} [hep-th].

\bibitem{Yang:2015bva}
  D.~L.~Yang and B.~M\"uller,
  {\it Collective flow of photons in strongly coupled gauge theories,}
  \arXivid{1507.04232} [hep-th].

\bibitem{Grozdanov:2014kva}
  S.~Grozdanov and A.~O.~Starinets,
  {\it On the universal identity in second order hydrodynamics,}
  \jhep {1503}{2015}{007},
  \arXivid{1412.5685} [hep-th].

\bibitem{Steineder:2012si}
    D.~Steineder, S.~A.~Stricker and A.~Vuorinen,
    {\it Holographic thermalization at intermediate coupling,}
    \prl {110}{2013}{101601}
    \arXivid{1209.0291} [hep-ph].

\bibitem{Stricker:2013lma}
  S.~A.~Stricker,
  {\it Holographic thermalization in $\Nfour$ super-Yang-Mills theory
  at finite coupling,}
  \epjc {74}{2014}{2727},
  \arXivid{1307.2736}.

\bibitem{Vuorinen:2013}
    D.~Steineder, S.~A.~Stricker, A.~Vuorinen,
    {\it Probing the pattern of holographic thermalization with photons,}
    \jhep {1307}{2013}{014},
    \arXivid{1304.3404}.

\bibitem{Paulos:2008tn}
  M.~F.~Paulos,
  {\it Higher derivative terms including the Ramond-Ramond five-form,}
  \jhep {0810}{2008}{047},
  \arXivid{0804.0763} [hep-th].

\bibitem{Myers:2008yi}
  R.~C.~Myers, M.~F.~Paulos and A.~Sinha,
  {\it Quantum corrections to $\eta/s$,}
  \prd {79}{2009}{041901},
  \arXivid{0806.2156} [hep-th].

\bibitem{Son:2002}
    D.~T.~Son, A.~O.~Starinets,
    {\it Minkowski-space correlators in AdS/CFT correspondence:
    recipe and applications,}
    \jhep {0209}{2002}{042},
    \hepth{0205051}.

\bibitem{Bak:2007fk}
  D.~Bak, A.~Karch and L.~G.~Yaffe,
  {\it Debye screening in strongly coupled $\Nfour$ supersymmetric Yang-Mills plasma,}
  \jhep {0708}{2007}{049},
  \arXivid{0705.0994} [hep-th].

\bibitem{Nunez:2003eq}
  A.~Nunez and A.~O.~Starinets,
  {\it AdS/CFT correspondence, quasinormal modes, and thermal correlators
  in $\Nfour$ SYM,}
  \prd {67}{2003}{124013},
  \hepth{0302026}.

\bibitem{Kovtun:2005}
    P.~K.~Kovtun, A.~O.~Starinets,
    {\it Quasinormal modes and holography,}
    \prd {72}{2005}{086009},
    \hepth{0506184}.

\bibitem{Boyd:2001}
    J.~P.~Boyd,
    {\it Chebyshev and Fourier spectral methods,}
    Dover (2001), 2nd ed.,
    \url{http://www-personal.umich.edu/~jpboyd/BOOK_Spectral2000.html}.


\bibitem{Benincasa}
  P.~Benincasa, A.~Buchel,
  {\it Transport properties of $\Nfour$ supersymmetric Yang-Mills
  theory at finite coupling,}
  \jhep {0601}{2006}{103},
  \hepth{0510041}.

\bibitem{Pawelczyk:1998pb}
  J.~Pawelczyk and S.~Theisen,
  {\it $AdS_5 \times S^5$ black hole metric at $O(\alpha'^3)$,}
  \jhep {9809}{1998}{010},
  \hepth{9808126}.

\end{thebibliography}
\end{document}